# In-plane antiferromagnetism in $Na_{0.5}CoO_2$ induced by $Sb_{Co}$ dopants

M. H. N. Assadi[1,*], S. Li[1], R. K. Zheng[2], S. P. Ringer[3,4] and A. B. Yu[5]
[1]*School of Materials Science and Engineering, UNSW Sydney, Sydney, NSW 2052, Australia.*
[2]*School of Physics, University of Sydney, Sydney, New South Wales 2006, Australia.*
[3]*Australian Centre for Microscopy & Microanalysis, University of Sydney, Sydney, New South Wales 2006, Australia.*
[4]*School of Aerospace, Mechanical and Mechatronic Engineering, University of Sydney, Sydney, New South Wales 2006, Australia.*
[5]*Department of Chemical Engineering, Monash University, Clayton, Vic. 3800, Australia.*
*e-mail address: h.assadi.2008@ieee.org*

$Na_xCoO_2$ has a fascinating and complex magnetic phase diagram that can be further manipulated by doping. Here, we investigated the effect of electron doping on the magnetic coupling among $Co^{4+}$ ions in $Na_xCoO_2$ using density functional theory based on Hartree-Fock hybrid functional. We found that electron doping through substitutional $Sb_{Co}$ dopants flip the in-plane ferromagnetic coupling among $Co^{4+}$ ions in the undoped compound to antiferromagnetic. Electron doping through interstitial $Cu_{Int}$, however, does not have a similar effect as $Cu_{Int}$ dopant leaves the compound ferromagnetic, just like the case of the undoped compound. The results demonstrate the critical dependence of magnetic phase in $Na_{0.5}CoO_2$ on the dopant and its incorporation site.

**Keywords:** Density functional theory, Dopant, Magnetism, Sodium cobaltate
**PACS:** 71.15.Mb, 75.20.−g, 71.27.+a, 75.47.Lx

## I. INTRODUCTION

Sodium cobaltate ($Na_xCoO_2$) is an interesting compound with rich magnetic and structural phase diagrams[1, 2] high thermoelectric figure of merit[3] and catalytic properties.[4] As shown in Fig. 1, the $Na_xCoO_2$ lattice is made of alternating Na layers and edge-sharing $CoO_2$ octahedra. The $CoO_2$ layer in $Na_xCoO_2$ has a high degree of electronic frustration and thus large spin entropy[5] and consequently large Seebeck coefficient.[6] Additionally, in $Na_xCoO_2$, phonons are strongly scattered by the $Na^+$ ions which are mobile and liquid-like at room temperature[7, 8] leading to low lattice thermal conductivity.[9-11]

The magnetic and structural phase diagram in $Na_xCoO_2$ is highly sensitive to Na concentration ($x$). In $Na_1CoO_2$, ($x = 1$), all Na ions occupied Na2 sites (Wyckoff *2c* position)[12, 13] and the parity between Na and Co ions creates a nonmagnetic electronic configuration $t_{2g}^6e_g^0$ for Co. For slightly lower Na content ($1 > x > 0.8$), the majority of the Na ions prefer to occupy the Na2 sites, however in order to reduce the electrostatic repulsion, some of the Na ions move to the Na1 site (Wyckoff *2b* position).[14] For instance, for $x = 85.71\%$, the Na2/Na is 5,[15] for $x = 83\%$, the Na2/Na1 ratio is 2.34,[12] and for $x = 81.25\%$, the Na2/Na1 ratio is 3.03.[16] In for $x = \sim 0.75$, the Na2/Na1 is further reduced to 2.00,[12, 14, 15] and Na ions are slightly off the oxygen axes along $c$ direction, resulting in a H3 structure.[17] Finally, for $Na_{0.5}CoO_2$ ($x = 0.5$), the Na2/Na1 ratio is 1.

In $Na_xCoO_2$, the Co ions adjacent to Na vacancies are in S= ½ ($t_{2g}^5e_g^0$) spin state and couple ferromagnetically within the $CoO_2$ layers for a wide value of Na concentrations.[7, 18-20] The magnetic coupling across the $CoO_2$ layers is, however, less straightforward. For instance, for $x$ values closer to 1, there is no general agreement on the nature of the out-of-plane coupling among Co ions; for $x = \sim 0.85$: both antiferromagnetic state with $T_N = 19.8$ K,[1, 21] and quasi-ferromagnetic state at $T_C = \sim 8$ K[22] have been reported. For $x = 0.75$, nonetheless, it has been confidently demonstrated that $Co^{4+}$ couple ferromagnetically within the same plane while Co ions on different $CoO_2$ layers couple antiferromagnetically, giving rise to A-type antiferromagnetic state.[14, 19, 23] Additionally, for $x = 0.5$, $Co^{4+}$ ions are couple ferromagnetically both in-plane and across layers.[14, 24] Regardless of the magnetic ground state, the origin of the out of plane coupling is generally attributed to second-neighbour superexchange interaction.[25]

In this work, we will demonstrate that electron doping in $Na_{0.5}CoO_2$ via the substitution of $Sb^{5+}$ for $Co^{3+}$ results in in-plane antiferromagnetic coupling among Co ions. Such peculiar magnetic interaction does not occur in the undoped $Na_xCoO_2$ compounds. The ferromagnetic coupling among Co ions is driven by a superexchange interaction between $Co^{4+}$ and $Co^{3+}$ ions through O 2p orbitals. Since the Co−O−Co angle is ~ 97° in $Na_xCoO_2$, ferromagnetism is favoured over antiferromagnetic interaction which requires wider Co−O−Co angle, usually wider than 120°.[26] We further show that electron doping via $Cu_{Int}$ which occupies an interstitial site in the Na layer does not induce in-plane antiferromagnetism.

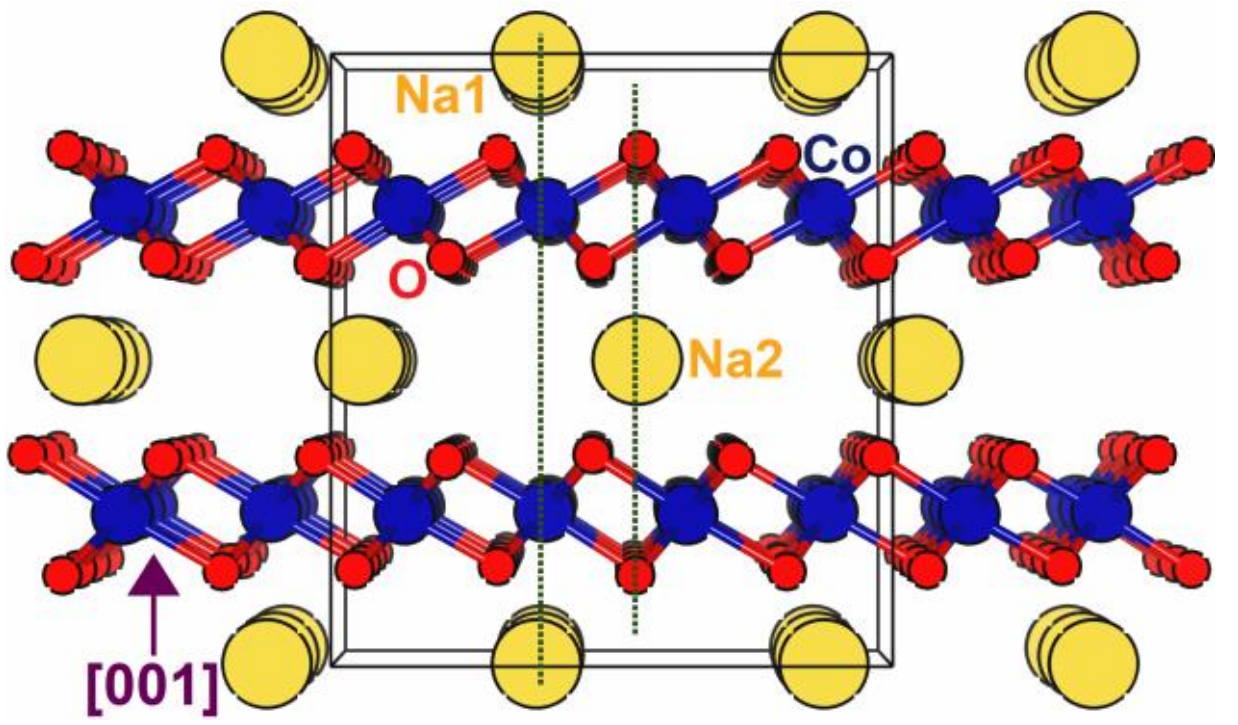

Fig. 1. A preview of the extended $Na_{0.5}CoO_2$ supercell of $P6_3/mmc$ space group. Co and O ions occupy the Wyckoff *2a* and *4f* sites of the hexagonal lattice structure respectively. In $Na_{0.5}CoO_2$ compound, half of the Na ions occupy *2b* (Na1) sites which share basal coordinates with Co and half of Na ions occupy *2c* (Na2) sites which share basal coordinates with O.

## II. COMPUTATIONAL SETTINGS

Spin-polarised density functional calculations, based on the projector augmented wave method,[27, 28] were performed with VASP package.[29, 30] For the final geometry optimisation, spin relaxation and the calculation of the density of states (DOS), the exchange-correlation functional was approximated with the hybrid method[31, 32] with a mixing parameter of 0.25. The hybrid functional reproduce the charge disproportionation[33, 34] of the Co ions more accurately. The energy cut-off was set to be 500 eV. Brillouin zone sampling for the supercells was carried out by choosing a 4 × 2 × 2 *k*-point set within Monkhorst-Pack scheme[35] with a grid spacing of ~ 0.05 $Å^{-1}$ between the *k* points. The lattice parameters of a fully optimised primitive unit cell of $Na_1CoO_2$ was found to be 2.87 Å for *a* and 10.90 Å for *c* which are in reasonable agreement with experimental lattice parameters[36] differing by −1.49 % and 0.07 % for *a* and *c* respectively. Then a $2a \times 4a \times 1c$ supercell was constructed for studying the doped compounds. Half of Na ions were removed according to the previously established pattern to create a $Na_8Co_{16}O_{32}$ supercell.[14] In our earlier works, we have already determined the most stable configuration of the Sb and Cu dopants in $Na_{0.5}CoO_2$ which were used in the present study.[37, 38]

## III. RESULTS AND DISCUSSION

In undoped $Na_xCoO_2$, the Co ions within a basal plane are generally arranged ferromagnetically while the out-of-plane coupling can be either ferromagnetic or antiferromagnetic depending on Na content (*x*). For $Na_{0.5}CoO_2$ earlier DFT and experimental studies have indicated a moderate ferromagnetic out-of-plane alignment among Co ions with a $T_c$ = 68 K.[14, 24] Now, let's first recall some of the aspects of Sb-doped $Na_xCoO_2$. First, it has been demonstrated that Sb dopants always replace a Co ion instead of being incorporated in the Na layer for $0.5 < x < 1$. That is proved by the lower formation energy ($E^f$) of $Sb_{Co}$ over other Sb doping configurations. Second, generally, the formation energy of $Sb_{Co}$ ever slightly decreases with decreasing *x*; $E^f(Na_1CoO_2{:}Sb_{Co})$ = 1.36 eV; $E^f(Na_{0.875}CoO_2{:}Sb_{Co})$ = 1.17 eV; $E^f(Na_{0.75}CoO_2{:}Sb_{Co})$ = 1.86 eV; $E^f(Na_{0.625}CoO_2{:}Sb_{Co})$ = 1.40 eV; $E^f(Na_{0.5}CoO_2{:}Sb_{Co})$ = 0.98 eV.[37, 39] This trend means that doping sodium cobaltite with lower *x* as is the case of $Na_{0.5}CoO_2$ is practically achievable.

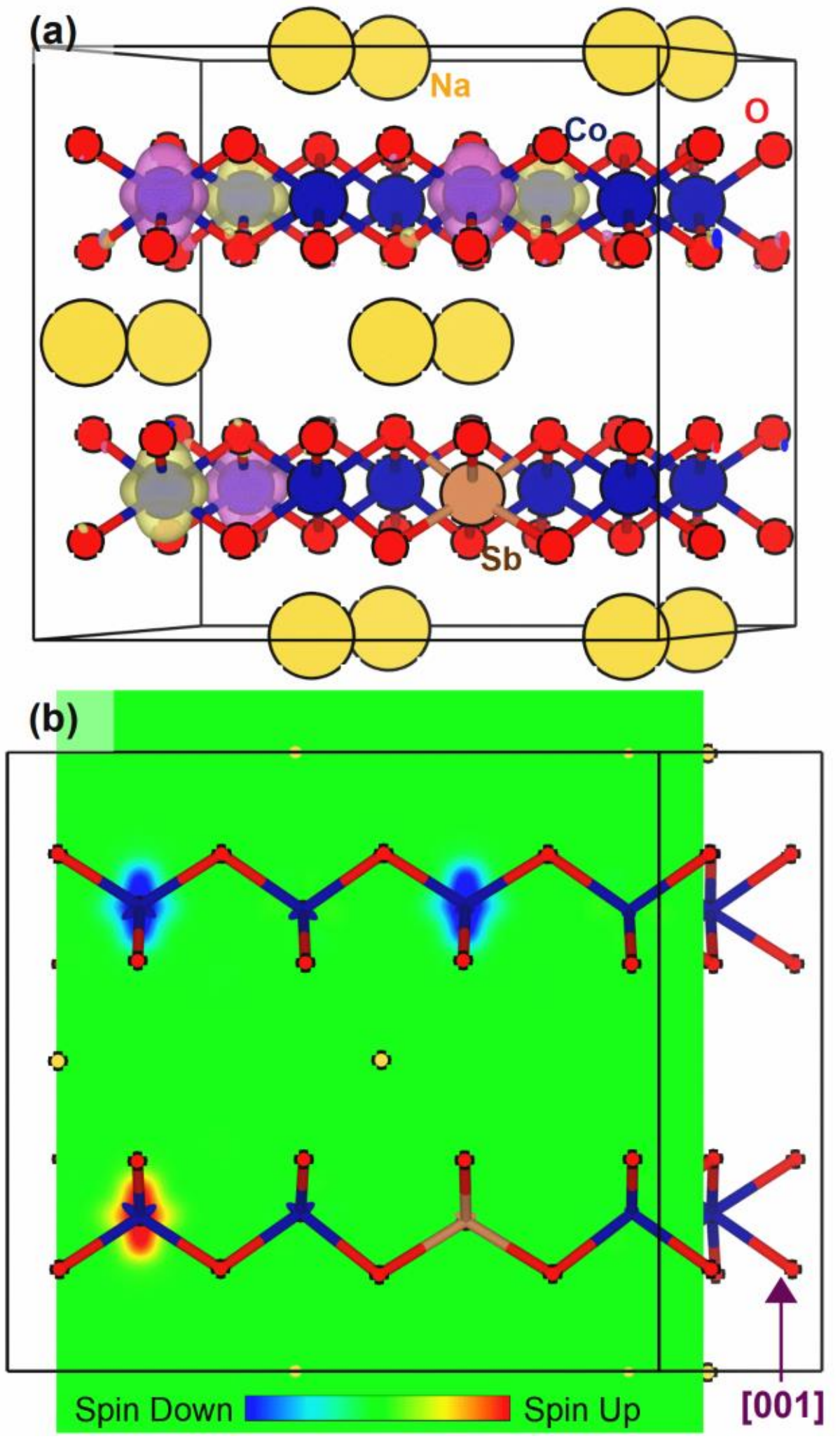

Fig. 2. (a) The $2a \times 4a \times 1c$ supercell used for simulating Sb-doped $Na_{0.5}CoO_2$ a1ong with spin density drawn at $\rho = \pm 0.015\ e/Å^3$. Purple and yellow iso-surface denote spin up and spin down respectively. (b) The spin density contours of a plane containing the $Sb_{Co}$ and Co ions along the *c* direction.

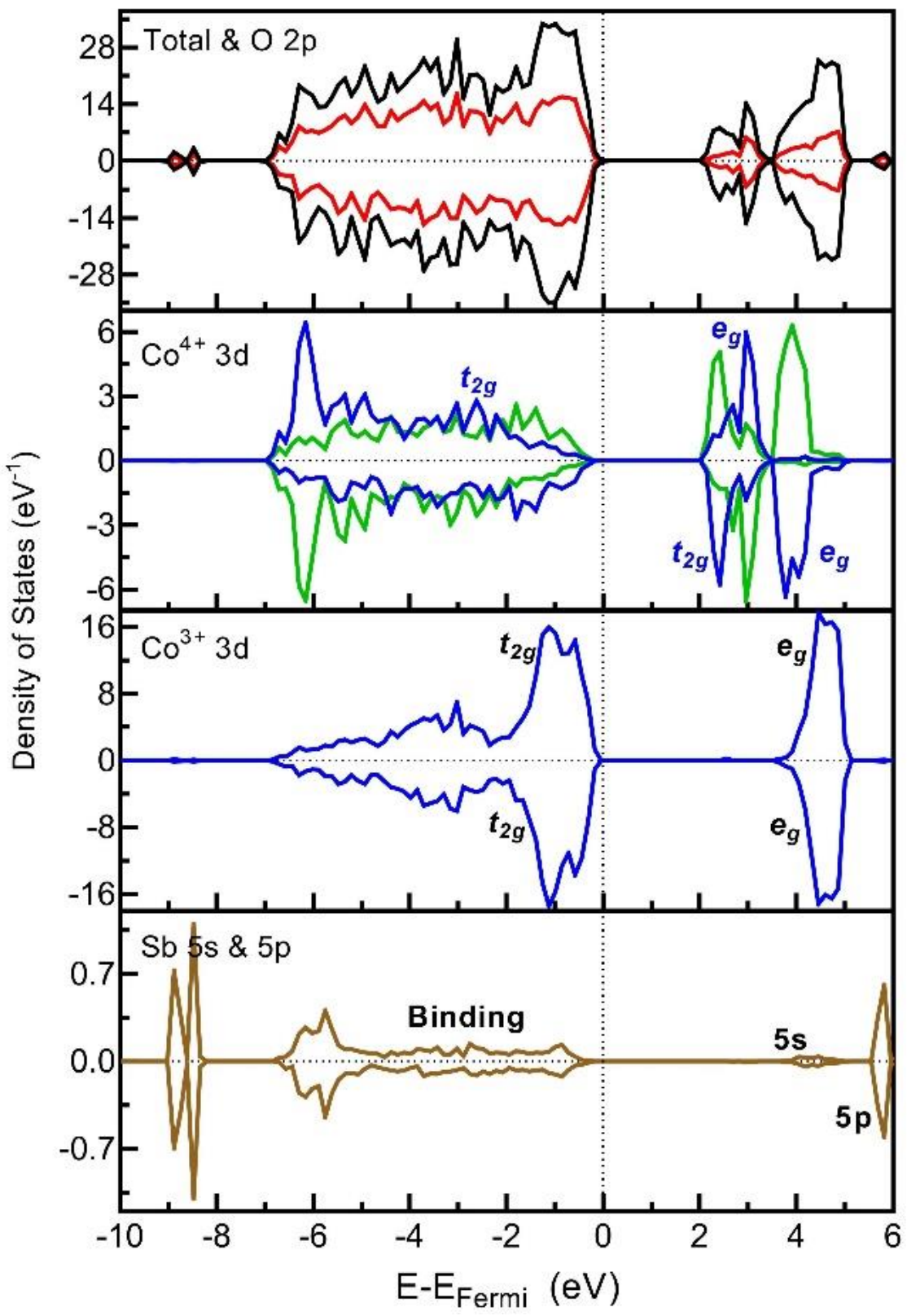

Fig. 3. Total and partial density of states of $Sb_{Co}$ doped $Na_{0.5}CoO_2$. The red and black lines in the top panel denote total and O 2p DOS respectively. The blue and green lines in the second panel denote the partial DOS of those $Co^{4+}$ ions that adapt net spin up and spin down directions respectively.

As shown in Fig. 2, the Sb doping in the $Na_{0.5}CoO_2$, however, fundamentally alters the magnetic interactions. First, there are only six spin-bearing Co ions ($Co^{4+}$, $t_{2g}^5e_g^0$) out of the 16 Co ions in the supercell. In the undoped $Na_xCoO_2$ compound, for every Na vacancy, a $Co^{3+}$ is reduced to $Co^{4+}$ to maintain charge neutrality. In other words, Na vacancies create holes that are borne on Co sites. So basically, before Sb doping, one would expect eight out of the 16 Co ions to be spin-bearing. The introduction of Sb dopant, however, neutralises two of those holes by having a high oxidation state of 5+. Sb oxidation state can be verified by examining the Bader charge on Sb ion, which was determined to be zero demonstrating that both 5s and 5p orbitals were empty. When $Sb^{5+}$ replaces a $Co^{3+}$, two extra electrons are introduced to the compounds that recombine with two holes that were originally created by Na vacancies leaving only six $Co^{4+}$ ions in the compound. Furthermore, the spin coupling between the neighbouring $Co^{4+}$ within the same $CoO_2$ layer changes from ferromagnetic in the undoped compound to antiferromagnetic in the $Sb_{Co}$ doped compound.

The partial and total density of states (DOS) of Sb-doped compound is given in Fig. 3. The fundamental bandgap in $Sb_{Co}$ doped $Na_{0.5}CoO_2$ is ~ 2.18 eV, which is rather closer to the gap in pristine $Na_1CoO_2$ which is ~ 2.3 eV[40, 41] and considerably wider than the bandgap of undoped $Na_{0.5}CoO_2$—estimated to be comfortably smaller than 0.2 eV.[14, 42, 43] As seen in the second panel of Fig. 3, the bandgap is determined by the position of the empty $t_{2g}$ states of $Co^{4+}$ which are located at ~ 2.315 eV above the occupied $Co^{4+}$ $t_{2g}$ states. The crystal field splitting between $2_{tg}$, and $e_g$ orbitals of the $Co^{3+}$ is ~ 3.68 eV which is larger than the fundamental bandgap and close to the similar gap in $Co_3O_4$.[44] Additionally, Sb's empty 5s states are located at the top of conduction band a ~ 4.2 eV hybridising with the empty $e_g$ states while Sb's empty 5p states are located higher than the conduction band at ~ 6 eV hybridising with the high lying Na's empty 3s states.

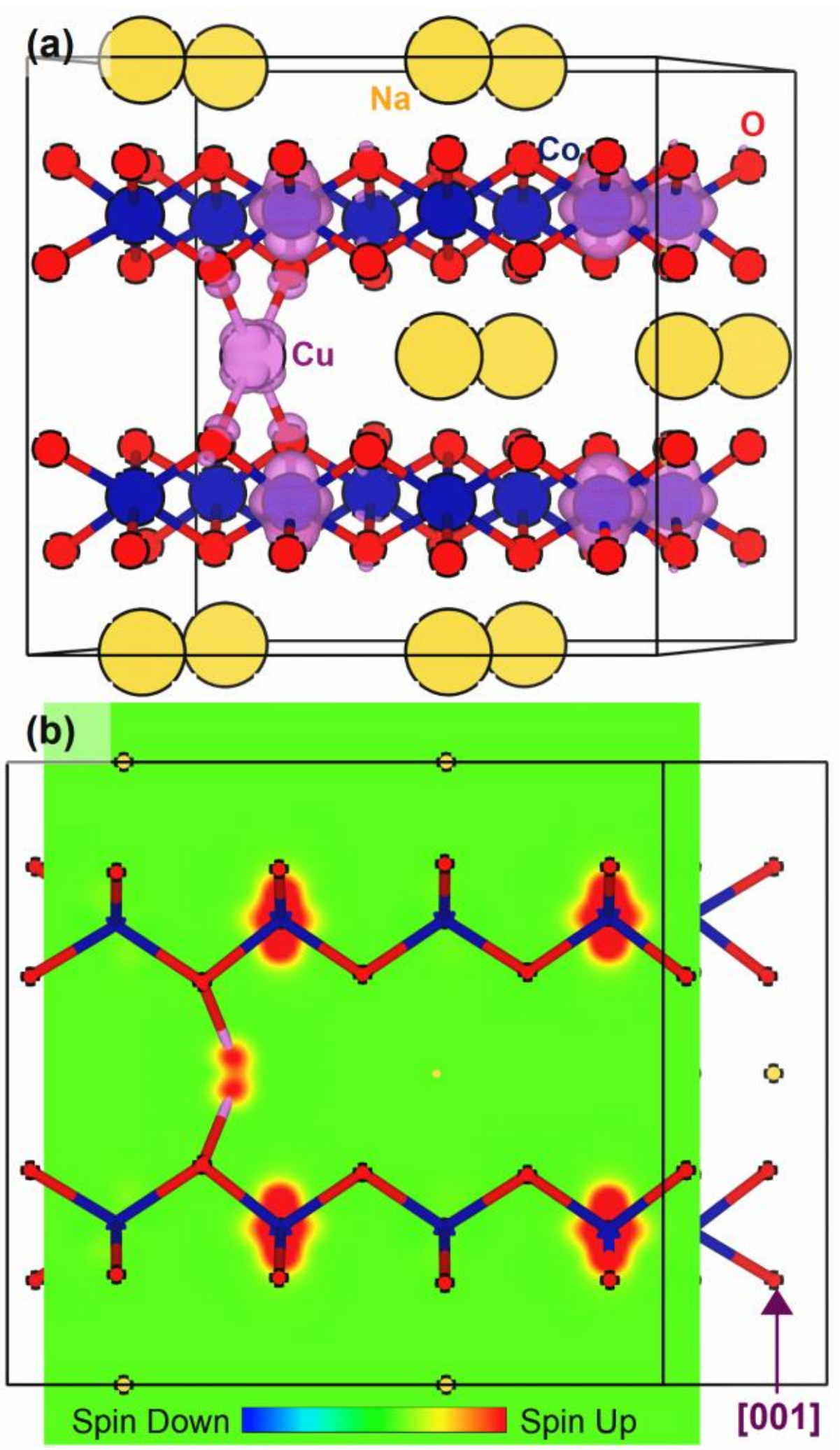

Fig. 4. (a) The $2a \times 4a \times 1c$ supercell used for simulating Cu doped $Na_{0.5}CoO_2$ a1ong with spin density drawn at $\rho = \pm 0.015$ e/Å$^3$. Purple and yellow iso-surface denote spin up and spin down respectively. (b) The spin density contours of a plane containing the $Cu_{Int}$ and Co ions along the $c$ direction.

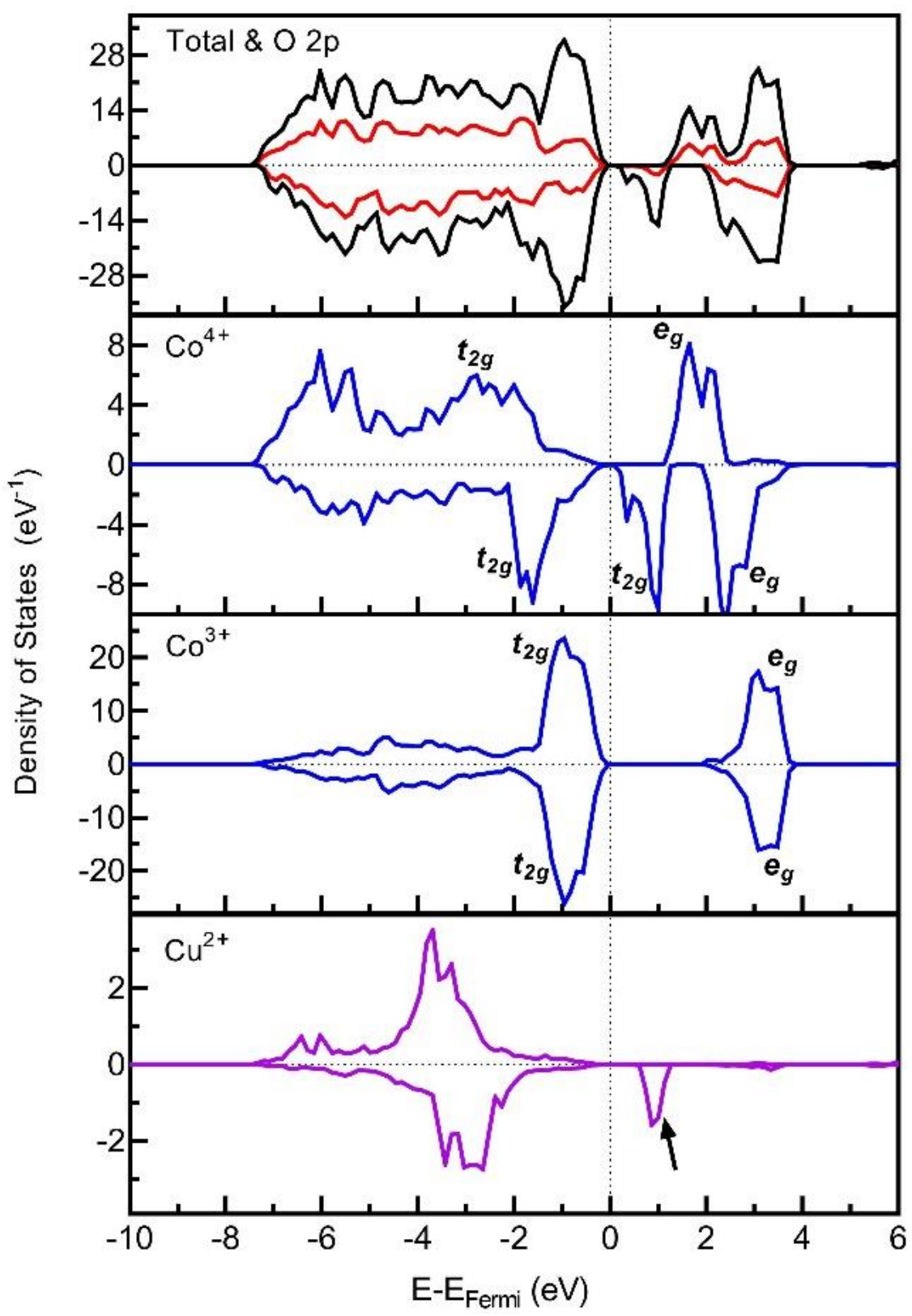

Fig. 5. Total and partial density of states of $Cu_{Int}$ doped $Na_{0.5}CoO_2$. The red and black lines in the top panel denote total and O 2p DOS respectively.

To see if the introduction of extra electrons on itself causes the in-plane antiferromagnetic coupling among $Co^{4+}$ ions, we examined the magnetic alignment in Cu doped $Na_{0.5}CoO_2$. As shown in Fig. 4, Cu is known to be incorporated interstitially in the Na layer of the $Na_xCoO_2$.[38] Positively charge $Cu_{Int}$ dopant also introduces electrons that can be recombined with the holes borne on Co sites. The spin density iso-surfaces in Fig. 4(a) indicates that only six Co ions bear spin. Following the same argument presented for the Sb-doped system, one infers that Cu has contributed two electrons to the compound. Fully oxidised $Cu^{2+}$ has a $d^9$ configuration, and consequently, Cu ion would bear spin which is evident from spin density visible on Cu site.

Unlike the case of Sb doping, the incorporation of $Cu_{Int}$ does not lead to in-plane antiferromagnetic coupling among the Co ions as seen in Fig. 4(a) and (b). Here, all $Co^{4+}$ ions are coupled ferromagnetically, both in-plane and out-of-plane. The ferromagnetic coupling is a remnant characteristic of the undoped compound which has not been altered by electron doping via $Cu_{Int}$. As seen in Fig. 5, Cu doping, moreover, does not affect the fundamental bandgap, which is 0.26 eV. The relatively meagre bang gap is maintained by two main mechanisms. First, the empty Cu d orbital (marked with an arrow) is located at the bottom of the valence band at ~ 0.87 eV. Moreover, the empty $t_{2g}$ orbitals of the $Co^{4+}$ ions strongly hybridise with the empty Cu d orbital and consequently are pulled down to lower energy ranges. The crystal field splitting between $Co^{3+}$'s filled $t_{2g}$ and empty $e_g$ states, 1.96 eV, is also slightly smaller than the case of the undoped compound.

Similar to the case of $Na_{0.5}CoO_2$, electron doping at the transition metal site has been found to induce in-plane antiferromagnetic coupling in $Sr_2FeMoO_6$ and $Ca_3Ru_2O_7$.[45, 46] This transition is driven by the diminishing of the kinetic-energy based superexchange mechanism as the number of $Co^{4+}$ ions decreases due to electron-hole recombination. Similarly, the transition from in-plane ferromagnetism to antiferromagnetism in $Na_{0.5}CoO_2$ critically depends on the doping site. We, therefore, conclude that the disruption of the hopping mechanism between $Co^{3+}$ and $Co^{4+}$ by $Sb_{Co}$ plays an important role in this transition.

## IV. CONCLUSIONS

In this work, using density functional theory with hybrid functional, the electronic and magnetic properties of $Sb_{Co}$ and $Cu_{Int}$ doped $Na_{0.5}CoO_2$ were studied. The main findings of this investigation can be summarised as follow: (1) The hybrid functional predicts that $Sb_{Co}$ adopts the highest oxidation state of 5+ (similar to the GGA+U functional) indicating the introduction of two electrons to the compound. (2) The $Sb_{Co}$ dopant widens the bandgap from approximately a ~ 0.2 of electron volt to 2.18 eV. (3) Sb dopants induce in-plane antiferromagnetic coupling among $Co^{4+}$ ions. This coupling is fundamentally different from the common in-plane ferromagnetic coupling in the undoped compounds. (4) Electron doping via $Cu_{Int}$ does not alter the ferromagnetic coupling among the $Co^{4+}$ ions of the undoped compound nor does it widen the bandgap. (5) The peculiar effect $Sb_{Co}$ might be attributed to the structural distortions caused by the larger $Sb^{5+}$ ionic radius.

## ACKNOWLEDGEMENTS

This work was supported by the Australian Research Council. The simulation facility was provided by Australian National Computational Infrastructure and Intersect Ltd.